\documentclass[preprintnumbers,amsmath,amssymb]{revtex4}

\usepackage{graphicx}
\usepackage{dcolumn}
\usepackage{bm}

\begin{document}

\title{Quantum secret sharing  between multiparty and multiparty  with four states}

\author{YAN Fengli$^{1,2}$, GAO Ting$^{2,3}$, LI Youcheng$^{1,2}$}

\affiliation {$^1$ College of Physics  and Information
Engineering, Hebei Normal
University, Shijiazhuang 050016, China\\
$^2$ CCAST (World Laboratory), P.O. Box 8730, Beijing 100080, China\\
$^3$ College of Mathematics and Information Science, Hebei Normal
University, Shijiazhuang 050016, China}

\date{\today}

\begin{abstract}
\textbf{Abstract}: An  protocol of quantum secret sharing between
multiparty and multiparty with four states is presented. We show
that this protocol can  make the Trojan horse attack  with a
multi-photon signal,   the fake-signal attack with EPR pairs, the
attack with single photons, and the attack with invisible photons to
be nullification. In addition, we also give the upper bounds of the
average success probabilities for dishonest agent eavesdropping
encryption using the fake-signal attack with any two-particle
entangled states.\\


\end{abstract}

\maketitle

 Recently we proposed  a quantum secret sharing (QSS) protocol between multiparty ($m$ members in group 1) and multiparty
($n$ members in group 2) using  a sequence of single photons
\cite{YGpra1}. In our protocol, all members in group 1 directly
encode their respective keys on the states of single photons via
unitary operations, then the last one (the $m$th member of group 1)
sends $1/n$ of the resulting qubits to each of group 2. After each
member of group 2 measures the photons, the two groups share the
secret messages. Unfortunately, Li {\it et al.} \cite {LiChangHwang}
pointed out that if the $m$th party of group 1 is dishonest, she can
obtain secret by substituting a sequence of single photons or a
sequence of Einstein-Podolsky-Rosen (EPR) pairs generated by himself
for the original photons without the detection of the other parties.
In this paper, we present a protocol of quantum secret sharing
between multiparty and multiparty with four states which is an
improvement of the one in \cite{YGpra1} and
 show that this improved protocol can prevent dishonest member from this kind
cheating. Moreover, the present protocol is also secure against the
Trojan horse attack with a multi-photon signal \cite {DYLLZG}, the
fake-signal attack with EPR pairs
 \cite {DLZZ} and the attack with invisible photons \cite {caiqingyu}.
We also give the upper bounds of the average success probabilities
for dishonest agent eavesdropping encryption using the fake-signal
attack with any two-particle entangled states.

\section{Review of the original  QSS protocol between multiparty and multiparty without entanglement}

Suppose that there are $m$ ($m\geq 2$) and $n$ ($n\geq 2$) members
in group 1 and group 2, respectively, and Alice 1, Alice 2,
$\cdots$, Alice $m$, and Bob 1, Bob 2, $\cdots$, Bob $n$ are their
respective all members. Group 1 wants quantum key sharing with group
2 such that neither  part of each group nor the combination of a
part of group 1 and a part of group 2 knows the key, but only all
members of each group can collaborate to determine what the string
(key) is. The original  QSS protocol between multiparty and
multiparty without entanglement \cite{YGpra1} is briefly summarized
as follows:

(M1) Alice 1 chooses two  random $nN$-bit strings $A_1$ and $B_1$.
She encodes   each bit $a_k^1$ of $A_1$ as
$|\psi_{a^1_kb^1_k}\rangle$, where $b^1_k$ is the corresponding bit
of $B_1$,
\begin{equation}\label{4qubits}
    \begin{array}{lll}
       |\psi_{00}\rangle &=& |0\rangle,\\
  |\psi_{10}\rangle &=& |1\rangle,\\
  |\psi_{01}\rangle &=& |+\rangle=\frac{|0\rangle+|1\rangle}{\sqrt{2}},\\
   |\psi_{11}\rangle &=& |-\rangle=\frac{|0\rangle-|1\rangle}{\sqrt{2}}.
     \end{array}
\end{equation}
 Then she sends the resulting $nN$-qubit state
$|\Psi^1\rangle=\otimes_{k=1}^{nN}|\psi_{a^1_kb^1_k}\rangle$ to
Alice 2.

(M2)  Alice 2 creates two random $nN$-bit strings $A_2$ and $B_2$.
She applies $\sigma_0=I=|0\rangle\langle0|+|1\rangle\langle1|$ or
$\sigma_1=\texttt{i}\sigma_y=|0\rangle\langle1|-|1\rangle\langle0|$
to each qubit $|\psi_{a^1_kb^1_k}\rangle$ of $nN$-qubit state
$|\Psi^1\rangle$ according to the corresponding bit of $A_2$ being 0
or 1, then she applies $I$ or
$H=\frac{1}{\sqrt{2}}(|0\rangle+|1\rangle)\langle0|+\frac{1}{\sqrt{2}}(|0\rangle-|1\rangle)\langle1|$
to each qubit of the resulting $nN$-qubit state depending on the
corresponding bit of $B_2$ being 0 or 1. After that, she sends Alice
3 the resulting $nN$-qubit state
$|\Psi^2\rangle=\otimes_{k=1}^{nN}|\psi_{a^2_kb^2_k}\rangle$.

(M3) Alice $i$ does likewise, $i=3, 4, \cdots, m$.  After Alice $m$
finishes her unitary operations on each single photon, she sends
$N$-qubit state
$\otimes_{j=0}^{N-1}|\psi_{a^m_{nj+l}b^m_{nj+l}}\rangle$ to Bob $l$,
$1\leq l\leq n$.

(M4) Bob 1, Bob 2, $\cdots$, Bob $n$ receive $N$ qubits,  and
announce this fact, respectively.

(M5) Alice 1, Alice 2, $\cdots$, and Alice $m$ publicly announce the
strings $B_1$, $B_2$, $\cdots$, and $B_m$, respectively.

(M6) Bob $l$ measures each of his qubit
$|\psi_{a^m_{nj+l}b^m_{nj+l}}\rangle$ in the $Z$ basis $\{|0\rangle,
|1\rangle\}$ (if $\oplus_{i=1}^m b_{nj+l}^i=0$) or in the $X$ basis
$\{|+\rangle=\frac {|0\rangle+|1\rangle}{\sqrt 2}, |-\rangle=\frac
{|0\rangle-|1\rangle}{\sqrt 2}\}$ (if $\oplus_{i=1}^m
b_{nj+l}^i=1$), $j=0, 1, \cdots, N-1$, $l=1, 2, \cdots, n$.

(M7) All Alices select randomly a subset  that will serve as a check
on Eve's interference, and tell all Bobs the bits they choose. In
the check procedure, all Alices and Bobs are required to broadcast
the values of their checked bits, and compare the XOR results of the
corresponding bits of checked bits of $A_1$, $A_2$, $\cdots$, $A_m$
and the values of the corresponding bits of Bob 1, Bob 2, $\cdots$,
and Bob $n$. If more than an acceptable number disagree, they abort
this round of operation and restart from first step.

(M8) The XOR results $\oplus^n_{l=1}(\oplus^m_{i=1}a_{nj_s+l}^i)$ of
Bob $l$'s corresponding bits $\oplus^m_{i=1} a_{nj_s+l}^i$ of the
remaining bits $nj_s+l$ of $\{\oplus^m_{i=1}
a_{nj+1}^i\}^{N-1}_{j=0}$,
  $\{\oplus^m_{i=1} a_{nj+2}^i\}^{N-1}_{j=0}$, $\cdots$, $\{\oplus^m_{i=1} a_{nj+n}^i\}^{N-1}_{j=0}$
  (or $\otimes_{j=0}^{N-1}|\psi_{a^m_{nj+1}b^m_{nj+1}}\rangle$,
$\otimes_{j=0}^{N-1}|\psi_{a^m_{nj+2}b^m_{nj+2}}\rangle$, $\cdots$,
 $\otimes_{j=0}^{N-1}|\psi_{a^m_{nj+n}b^m_{nj+n}}\rangle$) can be used as  key bits for secret sharing
  between all Alices and all Bobs.

 There exist  security flaws in this protocol. The details are as
 follows.

1. Alice 1 is a dishonest agent. Alice 1 prepares $nN$ EPR pairs in
the state $\otimes_{k=1}^{nN}|\Phi_k\rangle$, where
$|\Phi_k\rangle=\frac{1}{\sqrt{2}}(|00\rangle+|11\rangle)_{EA}$.
Alice 1 holds particle $E$ of each EPR pair and sends the sequence
$S_A$ of $\otimes_{k=1}^{nN}|\Phi_k\rangle$ to Alice 2. She does not
intercept $S_A$ until $S_A$ is sent  to Alice $i_1+1$ (or  Bobs).
 Note that after all operations of Alice 2, $\cdots$, and Alice
$i_1$ (or Alice $m$), $|\Phi_k\rangle$ is transformed into one of
the states
\begin{equation}\label{}
|\Phi^+\rangle, |\Psi^-\rangle,
\frac{1}{\sqrt{2}}(|\Phi^-\rangle+|\Psi^+\rangle),
\frac{1}{\sqrt{2}}(|\Phi^-\rangle-|\Psi^+\rangle),
\end{equation}
which are mutually orthogonal. Here
\begin{equation}\label{EPRpairs}
    \begin{array}{c}
       |\Phi^\pm\rangle  \equiv
       \frac{1}{\sqrt{2}}(|00\rangle\pm|11\rangle),
       |\Psi^\pm\rangle  \equiv
       \frac{1}{\sqrt{2}}(|01\rangle\pm|10\rangle).
    \end{array}
\end{equation}
 Since orthogonal states can be distinguish with certainty, Alice 1
 never makes a mistake and
 can steal  encoding information of Alice 2, $\cdots$, and Alice $i_1$ (or all Alices) freely and fully. In
 this way, Alice 1 can steal all Alices encryption fully and freely.
 After that, whether Alice 1 sends Bobs single photons or she sends Bobs EPR
 pairs, she and all Bobs share successfully the secret messages shared by all Alices and all Bobs.

2. Alice $i_0$ ($1\leq i_0\leq m$) is a dishonest agent. Alice $i_0$
generates $nN$ EPR pairs in the state
$\otimes_{k=1}^{nN}|\Phi_k\rangle$ as above Alice 1, where
$|\Phi_k\rangle=\frac{1}{\sqrt{2}}(|00\rangle+|11\rangle)_{EA}$.
When all Alices complete  their respective operations in (M1)-(M3),
Alice $i_0$ intercepts $|\Psi^m\rangle$, substitute $S_A$ for
$|\Psi^m\rangle$, and sends $N$ particles of   the sequence $S_A$ of
$\otimes_{k=1}^{nN}|\Phi_k\rangle$ to each Bob $l$. Alice $i_0$ can
obtain the original secret message by cheating without the detection
of other parties, which is the same as Alice $m$ using the attack
with EPR pairs in \cite{LiChangHwang}.

However, all the weaknesses  above can be avoided by the following
improved QSS protocol.

\section{the improvement of QSS protocol  between multiparty and multiparty  without entanglement}

 Now we give the  improved   quantum secret sharing scheme
as follows.

M1. Alice 1 begins with two random classical bit strings
$A_1=\{a^1_1, a^1_2, \cdots, a^1_{nN}\}$ and $B_1=\{b^1_1, b^1_2,
\cdots, b^1_{nN}\}$, where $a^1_k$ and $b^1_k$ are uniformly
chosen from $\{0, 1\}$. She then encodes these strings as a block
of $nN$ qubits,
\begin{eqnarray}\label{Alice1}
   |\Psi^1\rangle & = &\otimes_{k=1}^{nN}|\psi_{a^1_kb^1_k}\rangle\nonumber\\
     & = & \otimes_{j=0}^{N-1}|\psi_{a^1_{nj+1}b^1_{nj+1}}\rangle|\psi_{a^1_{nj+2}b^1_{nj+2}}\rangle\cdots
 |\psi_{a^1_{nj+n}b^1_{nj+n}}\rangle,\nonumber\\
\end{eqnarray}
where $a^1_k$ is the $k$th bit of $A_1$ (and similar for $B_1$),
each qubit $|\psi_{a^1_kb^1_k}\rangle$ is in one of the four states
in Eq.(\ref{4qubits}). The value of $b^1_k$ determines the basis. If
$b^1_k$ is 0 then $a^1_k$ is encoded in the $Z$ basis $\{|0\rangle,
|1\rangle\}$; if $b^1_k$ is 1 then $a^1_k$ is encoded in the $X$
basis $\{|+\rangle=\frac{|0\rangle+|1\rangle}{\sqrt{2}}$,
$|-\rangle=\frac{|0\rangle-|1\rangle}{\sqrt{2}}\}$. Because  the
four states are not all mutually orthogonal, therefore no
measurement can distinguish between  all of them with certainty.
Alice 1 then sends $|\Psi^1\rangle$ to Alice 2 over their public
quantum communication channel.

M2. When Alice $i$ receives  signals sent by Alice $i-1$, she
selects randomly a large subset of photons as the samples for
eavesdropping check. First, she use a special filter to prevent the
invisible photons from entering the operation system. Then she
splits each sample signal with a photon number splitter (PNS:
50/50), and  measures each signal in the measurement basis (MB) $Z$,
or $X$ at random \cite {DYLLZG}.
 Evidently if two photons in
one signal are detected, then Alice $i$ will abort the
communication. Moreover, she asks Alice 1, Alice 2, $\cdots$, Alice
$i-1$  to tell her their encoding information $a^t_s, b^t_s$ of the
samples in a random sequential order. After that,  she analyzes the
error rate $\varepsilon_s$ of the samples  she measured in MB $Z$ or
$X$ if $\oplus_{t=1}^{i-1}b_s^t=0$ or 1, respectively. In other
words, Alice $i$ analyzes the error rate $\varepsilon_s$ of the
samples she measured in the same basis with
$|\psi_{a^{i-1}_sb^{i-1}_s}\rangle$. If the error rate of the
samples   is  higher than a threshold $\varepsilon _r$,  then Alice
$i$  aborts the quantum communication. Otherwise she goes
 ahead. Here $s$ is the label of the sample chosen for eavesdropping check,  and
 $i=2,  3,  \cdots, m$.

 Clearly the test steps can avoid the Trojan horse attack  with a multi-photon
 signal \cite {DYLLZG}
 and the attack with invisible photons \cite {caiqingyu}.

M3.  Alice $i$  creates a quaternary  string $A_i=\{a^i_1, a^i_2,
\cdots, a^i_{nN+m^{i-1}}\}$ and a binary string $B_i=\{b^i_1, b^i_2,
\cdots, b^i_{nN+m^{i-1}}\}$, where $a^i_k$ and $b^i_k$ are uniformly
chosen from $\{0, 1, 2, 3\}$ and $\{0, 1\}$, respectively.
    For each $|\psi_{a^{i-1}_k b^{i-1}_k}\rangle$ of the $nN+m^{i-1}$ qubit state
 $|\Psi^{i-1}\rangle=\otimes_{i=1}^{nN+m^{i-1}}|\psi_{a^{i-1}_k b^{i-1}_k}\rangle$ , she performs the
    operation $\sigma_0$, $\sigma_1$, $\sigma_2$ or $\sigma_3$
 on it depending on the corresponding   $a^i_k$ of $A_i$ is 0, 1, 2
or 3, respectively.
 Simultaneously she has to operate the qubit with $I$ or a
Hadamard operator $H$ according to the bit $b^i_k$ in $B_i$ is 0 or
1, respectively. Here
\begin{eqnarray} &&\sigma_0=I=|0\rangle\langle 0|+|1\rangle\langle 1|,\nonumber\\
&&\sigma_1=\texttt{i}\sigma_y=-|1\rangle\langle 0|+|0\rangle\langle 1|,\nonumber\\
&&\sigma_2=\sigma_z=|0\rangle\langle 0|-|1\rangle\langle 1|,\nonumber\\
&&\sigma_3=\sigma_x=|0\rangle\langle 1|+|1\rangle\langle 0|,\nonumber\\
&&
H=\frac{1}{\sqrt{2}}(|0\rangle+|1\rangle)\langle0|+\frac{1}{\sqrt{2}}(|0\rangle-|1\rangle)\langle1|.
\end{eqnarray}
Alice $i$ making these unitary operations  is equal to the
encryption on the states of  single photons.
 The resulting state of this qubit is denoted by $|\psi_{a^i_k
b^i_k}\rangle$.  After that Alice $i$ inserts randomly $m^i-m^{i-1}$
decoy single photons  into $nN+m^{i-1}$ photons encoded by her,
where each of  the decoy single photons is   randomly in one of the
states in  Eq.({\ref{4qubits}}). Here $i=2, 3, \cdots, m$.

M4. Alice $i$ sends the photons ($nN+m^i$ qubits in the state
$|\Psi^i\rangle=\otimes_{i=1}^{nN+m^i}|\psi_{a^i_k b^i_k}\rangle$)
to Alice $i+1$ ($i=2, 3, \cdots, m-1$). After Alice $m$ completes
her encoding operations and randomly inserts $m^m-m^{m-1}$ decoy
single photons, she sends $nN+m^m$ photons to Bob 1, Bob 2,
$\cdots$, Bob $n$ in a sequential
 order.

M5. When all Bob 1, Bob 2, $\cdots$, and Bob $n$ have  received
their respective strings of  qubits, each of them first randomly and
independently chooses sufficient  samples to make measurement in MB
$Z$, or $X$ randomly. Then they
 ask Alice 1, Alice 2, $\cdots$, and Alice $m$ to announce publicly the $a^i_s$, $b^i_s$,  of the samples in a
random sequential order. Here $i=1,2,\cdots, m$. After that all Bobs
publish their measurement outcomes and the MBs.    If the error rate
of the samples is  higher than a threshold,
 then they abort the quantum communication. Otherwise they go to the next step.

 M6. All Alices ask all Bobs to delete the  decoy qubits which are  not chosen for eavesdropping check.
 All members  in group 1 publicly announce the strings $B_1, B_2, \cdots, B_m$ in a random sequential order.

 M7. Bob $l$ ($l=1, 2, \cdots, n$) measures each of their qubits with the MB $Z$ or $X$ according to the XOR
  results of the corresponding bits in the strings
  $B_1, B_2, \cdots, B_m$. That is, Bob $l$
measures $|\psi_{a^m_{nj+l}b^m_{nj+l}}\rangle$ in the $Z$ basis  if
$\oplus_{i=1}^m b_{nj+l}^i=0$ or in the $X$ basis  if
$\oplus_{i=1}^m b_{nj+l}^i=1$.
  From  his measurement on $|\psi_{a^m_{nj+l}b^m_{nj+l}}\rangle$, he obtains the outcome
  $d_{nj+l}$, which is 0 or 1, corresponding to the +1 and $-1$   eigenstates  of  $\sigma_x$ and $\sigma_z$.

 M8. All members in group 1 complete the error rate analysis of the transmission between the two groups. To this
 end, all Alices require each of the member in group 2 to publish a subset of the measurement results chosen randomly, and
 analyze the error rates of the samples.
  If the channel is secure, the XOR results of measurement outcomes
   of Bob 1, Bob 2, $\cdots$, and   Bob $n$'s corresponding
 bits can be used as key bits for secret sharing, otherwise they discard the results obtained and re-try the
 quantum communication from the beginning.

\section{Security}

Obviously,  the test in M2  can make the attack with invisible
photons \cite{caiqingyu} and the Trojan horse attack \cite{DYLLZG}
to be nullification. The checking procedure in M5 can avoid the
attack with single photons and the attack with EPR pairs
\cite{LiChangHwang}.  The fake-signal attack with EPR pairs
\cite{DLZZ} can be detected by the operations in M3 and the check in
M2 and  M5. The security of this present QSS protocol  against the
attacks in \cite{LiChangHwang, DLZZ} are discussed as follows.

\subsection{The security against the attack with single photons and the attack with EPR pairs}

Now, we show that  the attack stated in Ref. \cite {LiChangHwang}
 is easily detected in the present  quantum secret sharing.

Apparently, Bob's measurements on the checked samples collapse
them
 into the states ( in Eq.(\ref{4qubits}) ) of  single photons. In other words, all
Bob's measurements on the checked samples remove the entanglements
between checked photons and other eavesdropping particles, which
corresponds to that the attacker Alice $i_0$ sends Bobs single
quantum states in Eq.(\ref{4qubits}) whether in the attack  with EPR
pairs or in the attack with single photons \cite{LiChangHwang}.
Thus, all Alices and all Bobs can find out the attacker in the
attack with single photons via all Alices' publishing their
respective encoding information in a random sequential order. Since
Bob $l$ asks Alices to announce $a^i_s$ and $b^i_s$ in a random
sequential order, the attacker will not be the last one to answer
the Bob $l$'s enquiry with a probability $\frac {m-1}{m}$. If there
is an Alice to be asked after the attacker Alice $i_0$, Alice $i_0$
can only guess $a^{i_0}_s$ and $b^{i_0}_s$ to answer the inquiry as
she
 can not distinguish the quantum state intercepted by her with
certainty. It is easy to derive  that the error rate of the samples
that Bobs measured in MB $Z$ or $X$ according to
$\oplus^m_{i=1}b^i_s$=0 or 1 is more than $\frac {m-1}{2m}$.

\subsection{The security against the fake-signal attack with EPR pairs}

The present  QSS protocol is secure against any dishonest agent
eavesdropping secret using the fake-signal attack with EPR pairs
\cite{DLZZ}. We shall actually show that this protocol is secure
against more general attack, the fake-signal attack with any
two-particle entangled states (general EPR pairs).  Next let us see
how this works in detail.

Suppose that the malicious party Alice $i_0$ who may   be any
dishonest one of Alices, generates $nN+m^{i_0}$ general EPR pairs in
the state $\otimes_{k=1}^{nN+m^{i_0}}|\phi_k\rangle$, where
\begin{equation}\label{}
   |\phi_k\rangle=|\phi\rangle=|0\rangle_A|\alpha\rangle_E+|1\rangle_A|\beta\rangle_E,
\end{equation}
 $|\alpha\rangle_E$ and $|\beta\rangle_E$  are unnormalized
  states of the $S$-level ($S\geq 2$) particle $E$. Note that when
\begin{equation}\label{orthogonal}
           \langle\alpha|\beta\rangle = \langle\beta|\alpha\rangle=0,
  \langle\alpha|\alpha\rangle = \langle\beta|\beta\rangle=\frac{1}{2},
 \end{equation}
$|\phi_k\rangle=|\phi\rangle$ is an EPR pair, that is, EPR pair is
   the special case of
  $|\phi\rangle$.
Alice $i_0$  replaces each of the original single photons
$|\Psi^{i_0}\rangle$ with a fake signal, the general EPR pair
$|\phi_k\rangle$ and sends the sequence $S_A$ of $nN+m^{i_0}$
particles $A$ in $\otimes_{k=1}^{nN+m^{i_0}}|\phi_k\rangle$  to
Alice $i_0+1$ while each particle $E$ in
$\otimes_{k=1}^{nN+m^{i_0}}|\phi_k\rangle$ is held by herself.   If
$|\phi\rangle$ is not a two-particle maximally entangled state, then
the cheating of Alice $i_0$ can be found by Alice $i_0+1$ in M2,
because Alice $i_0$ can not distinguish between $|\alpha\rangle$ and
$|\beta\rangle$ and between $|\alpha\rangle+|\beta\rangle$ and
$|\alpha\rangle-|\beta\rangle$ perfectly,  it is a certain that
Alice $i_0$ makes mistakes. Next we only assume that $|\phi\rangle$
is an EPR pair.  Alice $i_0+1$ in the step M2 cannot detect this
cheating as Alice $i_0$ is able to produce no errors in the results
if Alice $i_0$ is asked to announce her encryption $a_s^{i_0}$ and
$b_s^{i_0}$ of the samples after Alice 1, $\cdots$, Alice $i_0-1$.
But if Alice $i_0$ is not the last to announce her encoding
information, then her cheating introduces errors and can be found
out by Alice $i_0+1$ in M2 without fail. However, when the dishonest
Alice $i_0$ is Alice 1,  this cheating of her cannot be found out by
Alice 2 as it does not introduce errors in the results.

 Alice 1 intercepts $S_A$ while it was sent to Alice $i_1$
($2<i_1\leq m$) or Bobs (evidently, if Alice 1 never intercepts
$S_A$, then she can not obtain any information, though this kind of
eavesdropping can not be found in the eavesdropping check. So it
does not make any sense for Alice 1 to do this kind of eavesdropping
). Note that the result of the encryptions by Alice 2, $\cdots$,
Alice $i_1-1$ or Alice $m$ in step M3 is equivalent to one Alice's
operations in M3 (that is, one Alice performs one of operations
$\sigma_0$, $\sigma_1$, $\sigma_2$, $\sigma_3$, $H\sigma_0$,
$H\sigma_1$, $H\sigma_2$, $H\sigma_3$ on each particle $A$). This is
because of $H\sigma_1H=-\sigma_1$, $H\sigma_2H=\sigma_3$,
$H\sigma_3H=\sigma_2$, $H^2=\sigma_u^2=I$,
$\sigma_u\sigma_v=-\sigma_v\sigma_u$ for $u,v=1,2,3$.

The object of Eve is to get all Alices' encoding information   $A_i$
and $B_i$.
  Therefore, Alice 1 must manage to  distinguish the eight operations $\sigma_0$, $\sigma_1$, $\sigma_2$, $\sigma_3$,
$H\sigma_0$, $H\sigma_1$,  $H\sigma_2$, and $H\sigma_3$ with which
$|\phi\rangle$ is transformed  into one of eight states
\begin{eqnarray}\label{8states}
 &&|\varphi_1\rangle =\sigma_0|\phi\rangle=|0\rangle|\alpha\rangle+|1\rangle|\beta\rangle,\nonumber\\
 &&|\varphi_2\rangle =\sigma_1|\phi\rangle=-|1\rangle|\alpha\rangle+|0\rangle|\beta\rangle,\nonumber\\
 &&|\varphi_3\rangle =\sigma_2|\phi\rangle=|0\rangle|\alpha\rangle-|1\rangle|\beta\rangle,\nonumber\\
 &&|\varphi_4\rangle =\sigma_3|\phi\rangle\nonumber
 = |1\rangle|\alpha\rangle+|0\rangle|\beta\rangle,\nonumber\\
 &&|\varphi_5\rangle=H\sigma_0|\phi\rangle =\frac{1}{\sqrt 2}(|0\rangle+|1\rangle)|\alpha\rangle+\frac{1}{\sqrt
 2}(|0\rangle-|1\rangle)|\beta\rangle,\nonumber\\
  &&|\varphi_6\rangle=H\sigma_1|\phi\rangle= -\frac{1}{\sqrt 2}(|0\rangle-|1\rangle)|\alpha\rangle
 +\frac{1}{\sqrt 2}(|0\rangle+|1\rangle)|\beta\rangle,\nonumber\\
  &&|\varphi_7\rangle=H\sigma_2|\phi\rangle=\frac{1}{\sqrt 2}(|0\rangle+|1\rangle)|\alpha\rangle
 -\frac{1}{\sqrt 2}(|0\rangle-|1\rangle)|\beta\rangle,\nonumber\\
  &&|\varphi_8\rangle=H\sigma_3|\phi\rangle=\frac{1}{\sqrt 2}(|0\rangle-|1\rangle)|\alpha\rangle
 +\frac{1}{\sqrt 2}(|0\rangle+|1\rangle)|\beta\rangle,
 \end{eqnarray}
where $|\phi\rangle$ is an EPR pair (i.e. $|\alpha\rangle$ and
$|\beta\rangle$ satisfy Eq.(\ref{orthogonal})). In other words, she
is to distinguish the eight states.
 The eight states in Eq.(\ref{8states})
belongs to the subspace $W$ of the entire Hilbert space
$\mathcal{H}$ spanned by $|0\rangle|\alpha\rangle$,
$|0\rangle|\beta\rangle$, $|1\rangle|\alpha\rangle$ and
$|1\rangle|\beta\rangle$. Obviously, the dimension $\texttt{dim}W$
of subspace $W$ is less than or equal to 4. It implies that it is
impossible for all eight states in Eq.(\ref{8states}) being mutually
orthogonal, so these eight states can not be reliably distinguished
\cite{NC}. Thus Alice 1's eavesdropping can be detected in M2 by
Alice $i_1$ or M5 by all Bobs. Therefore the fake-signal attack with
EPR pairs can not work for the present quantum secret sharing
protocol.

On the other hand, in the present protocol, Alices  insert randomly
decoy single photons into the signal photons in M3. The
eavesdropping check on the decoy single photons is the same as that
on the signal photons. That is, first Alice $i$ measures (or all
Bobs measure) each decoy single photon in MB $Z$, or $X$ at random,
then she  asks (they ask) Alice 1, Alice 2, $\cdots$, Alice $i-1$
(all Alices) to tell her (them) their encoding information $a^t_s,
b^t_s$ of the samples in a random sequential order.  Note that there
is at least one honest agent  in one communication group. Therefore
the dishonest agent can be found by the eavesdropping checks on the
decoy photons by the honest agents. The principle of the checking
procedures is the same as that in BB84 quantum key distribution
protocol \cite{NC}.

Remark 1. If only Alice 2 is honest in group 1, and all Alices
except  Alice 2 collude,  the cheating of Alice 1 can be found out
by all Bobs in the step M5.

Remark 2. Alice 1 can not  unambiguously discriminate four sets
\begin{eqnarray}\label{4sets}
\{|\varphi_{11}\rangle, |\varphi_{12}\rangle\},
\{|\varphi_{21}\rangle,
|\varphi_{22}\rangle\},\{|\varphi_{31}\rangle,
|\varphi_{32}\rangle\},\{|\varphi_{41}\rangle,
|\varphi_{42}\rangle\},\nonumber\\
\end{eqnarray}
where
\begin{eqnarray} &&|\varphi_{11}\rangle=|\varphi_1\rangle,
|\varphi_{12}\rangle=|\varphi_5\rangle,|\varphi_{21}\rangle=|\varphi_2\rangle,
|\varphi_{22}\rangle=|\varphi_6\rangle,\nonumber\\
&& |\varphi_{31}\rangle=|\varphi_3\rangle,
|\varphi_{32}\rangle=|\varphi_7\rangle,
|\varphi_{41}\rangle=|\varphi_4\rangle,
|\varphi_{42}\rangle=|\varphi_8\rangle.\nonumber\\
\end{eqnarray}
This will be given later. Even  if Alice 1 can  correctly classify
the four sets in (\ref{4sets}), she will be detected in the step M2
by Alice $i_1$ or in the step M5 by all Bobs  as she produces errors
when answering Alice $i_1$' or all Bobs' enquiry.

Remark 3. It is not necessary for Alice $i$ ($2\leq i\leq m$)
performing four operations $\sigma_0$, $\sigma_1$, $\sigma_2$ and
$\sigma_3$ in M3. In fact, Alice $i$  only needs to use three
operations either $\sigma_0$, $\sigma_1$, $\sigma_2$ or $\sigma_0$,
$\sigma_1$, $\sigma_3$.

 Next we will calculate the upper bounds of the average success probabilities  of the two cases, one is to distinguish the eight states in
 Eq.(\ref{8states}), the other is to classify the four sets in
 (\ref{4sets}), where $|\phi\rangle$ is a  two-particle  entangled state.

Case I. The upper bound of the average success probability
distinguishing the eight states in Eq.(\ref{8states}). Let
\begin{eqnarray}
&&x=\langle\alpha|\beta\rangle+\langle\beta|\alpha\rangle,\nonumber\\
&&q=\frac{1}{\texttt{i}}(\langle\alpha|\beta\rangle-\langle\beta|\alpha\rangle),\nonumber\\
&&z=\langle\alpha|\alpha\rangle,\nonumber\\
&&t=\langle\beta|\beta\rangle,\nonumber\\
&&z+t=1,
\end{eqnarray}
 we have
\begin{eqnarray}
&&\sum_{\substack{i,j=1 \\ i\neq j}}^8 |\langle\varphi_i|\varphi_j\rangle|\nonumber\\
= && 8|q|+8|z-t|+8|x|+\frac {8}{\sqrt 2}|z-t+x|\nonumber\\
&&+\frac {8}{\sqrt 2}|z-t-x|+\frac {16}{\sqrt 2}\sqrt {1+q^2}.
\end{eqnarray}
It follows that minimum of $\sum_{\substack{i,j=1 \\ i\neq j}}^8
|\langle\varphi_i|\varphi_j\rangle|$ can be realized if
\begin{eqnarray}
q=0, z=t=\frac {1}{2}, x=0.
\end{eqnarray}
It means that the minimum of $\sum_{\substack{i,j=1 \\ i\neq j}}^8
|\langle\varphi_i|\varphi_j\rangle|$ occurs when $|\phi_k\rangle$ is
an EPR pair.

The minimum  of $\sum_{\substack{i,j=1 \\ i\neq j}}^8
|\langle\varphi_i|\varphi_j\rangle|$ is
\begin{equation}
 \textrm{Min}{\sum_{\substack{i,j=1 \\ i\neq j}}^8
|\langle\varphi_i|\varphi_j\rangle|}=\frac {16}{\sqrt 2}.
\end{equation}
The average success probability $P_1$ \cite{ZFSY} for unambiguous
identification of the eight states in Eq.(\ref{8states}) satisfies
\begin{eqnarray}\label{p1}
P_1\leq && 1-\frac {1}{M-1}\sum_{\substack{i,j=1 \\ i\neq j}}^8
\sqrt
{p_ip_j}|\langle\varphi_i|\varphi_j\rangle|\nonumber\\
=&&1-\frac{1}{56}\times\frac {16}{\sqrt 2}\nonumber\\
=&&1-\frac {\sqrt 2}{7}<1,
\end{eqnarray}
where $M=8$ is the number of the states to be distinguished,
$p_i=\frac{1}{8}$ and $p_j=\frac{1}{8}$ are the prior probabilities
of $|\varphi_i\rangle$ and $|\varphi_j\rangle$, respectively.

Case II. The upper bound of the average success probability
classifying the four sets in Eq.(\ref{4sets}). By calculation, we
obtain that
\begin{eqnarray}
&&\sum_{\substack{i,j=1 \\ i\neq j}}^4\sum_{k,l=1}^2\sqrt {\frac
{\eta_{ik}\eta_{jl}}{
{(N-m_i)(N-m_j)}}}|\langle\varphi_{ik}|\varphi_{jl}\rangle|\nonumber\\
=&&\sum_{\substack{i,j=1 \\ i\neq j}}^4\sum_{k,l=1}^2\sqrt {\frac
{\frac{1}{8}\times\frac{1}{8}}{
{(8-2)\times(8-2)}}}|\langle\varphi_{ik}|\varphi_{jl}\rangle|\nonumber\\
=&&\frac {1}{12}(2|y|+\frac {1}{\sqrt 2}|z-t-x|+2|z-t|\nonumber\\
&&+\frac {4}{\sqrt 2}\sqrt {1+q^2}+2|x|+\frac {1}{\sqrt
2}|z-t+x|).\end{eqnarray} Here $N=8$ is the total number of the
states in the four classified sets in (\ref{4sets}), $m_i$  is the
number of the states of the $i$-th set, and $\eta_{ik}$ is the prior
probability of $|\varphi_{ik}\rangle$.

 A little thought shows that
\begin{equation}
\sum_{\substack{i,j=1 \\ i\neq j}}^4\sum_{k,l=1}^2\sqrt {\frac
{\eta_{ik}\eta_{jl}}{
{(N-m_i)(N-m_j)}}}|\langle\varphi_{ik}|\varphi_{jl}\rangle|\geq
\frac{\sqrt{2}}{6},
\end{equation}
and the equality holds if
\begin{equation}
x=0, z=t=\frac {1}{2}, q=0,\end{equation}
 that is, the minimum of $\sum_{\substack{i,j=1 \\
i\neq j}}^4\sum_{k,l=1}^2\sqrt {\frac {\eta_{ik}\eta_{jl}}{
{(N-m_i)(N-m_j)}}}|\langle\varphi_{ik}|\varphi_{jl}\rangle|$ reaches
when $|\phi\rangle$ is an EPR pair.

 The average success
probability $P_2$ of conclusive quantum states sets classification
\cite{WY} is
\begin{eqnarray}\label{p2}
P_2&&\leq 1-\sum_{\substack{i,j=1 \\ i\neq j}}^4\sum_{k,l=1}^2\sqrt
{\frac {\eta_{ik}\eta_{jl}}{
{(N-m_i)(N-m_j)}}}|\langle\varphi_{ik}|\varphi_{jl}\rangle|\nonumber\\
&&=1-\frac {\sqrt 2}{6}<1.
\end{eqnarray}

From (\ref{p1}) and (\ref{p2}), no matter what kind strategy the
malicious Alice $i_0$ use, she will disturb the quantum system, make
mistakes,  and therefore  can be detected in M2, M5 and M8.
Consequently, the present  QSS protocol is secure  not only against
the fake-signal attacking with EPR pairs but also against the
fake-signal attacking with any two-particle entangled
states---general EPR pairs.

This present QSS protocol is  an  improvement of original QSS
protocol \cite {YGpra1}. In the present QSS protocol, we add the
special filters, photon number splitters, single-photon detectors,
the eavesdropping check of each member Alice $i$ ($i=2,3,\cdots, m$)
in group 1, decoy single photons, two operations $\sigma_x$ and
$\sigma_z$, and the random measurements of all Bobs on their
respective qubits chosen at random.
 It is the  improvement that  makes the attacks
pointed in \cite {LiChangHwang, DYLLZG, DLZZ, caiqingyu} to be of no
effect (to be nullification). The principle of the checking
procedures is the same as that in BB84 quantum key distribution
protocol. So the transmission of qubits between
 authorized members in the two groups is secure. That is, the
 present  QSS protocol between multiparty and multiparty with four states is secure.

\acknowledgments This work is supported by the National Natural
Science Foundation of China under Grant No:10671054 and Hebei
Natural Science Foundation of China under Grant No: A2004000141 and
A2005000140.


\begin{thebibliography}{999}
\bibitem{YGpra1}  Yan F L, Gao T. Quantum secret sharing between multiparty and multiparty without entanglement. Phys  Rev  A, 2005, 72(1): 012304
\bibitem{LiChangHwang}  Li C M,  Chang C C,   Hwang T. Comment on "Quantum secret sharing between multiparty and multiparty without entanglement"
, Phys Rev A, 2006, 73(1): 016301
\bibitem{DYLLZG}  Deng F G,  Yan F L,  Li X H,  et al. Addendum to "Quantum secret sharing between multiparty and multiparty without entanglement"
, arXiv: quant-ph/0508171
\bibitem{DLZZ}  Deng F G,  Li X H,  Zhou H Y,   et al. Erratum: Improving the security of multiparty quantum secret sharing against Trojan horse attack [Phys. Rev. A 72, 044302
(2005)].  Phys Rev A, 2006, 73(4): 049901
\bibitem{caiqingyu}  Cai Q Y. Eavesdropping on the two-way quantum communication protocols with invisible photons. Phys Lett A, 2006, 351(1): 23-25
\bibitem{NC}  Nielsen  M A,  Chuang I L. Quantum Computation and Quantum Information.
          Cambridge: Cambridge University Press, 2000
\bibitem{ZFSY}  Zhang S,  Feng Y,  Sun X,  et al. Upper bound for the success probability of unambiguous discrimination among quantum states
. Phys  Rev  A, 2001, 64(6): 062103
\bibitem{WY}  Wang  M Y, Yan F L. Conclusive quantum state classification, arXiv: quant-ph/0605127
\end{thebibliography}
\end{document}